\documentclass[aps,prl,twocolumn,showpacs,superscriptaddress,preprintnumbersfleqn,floatfix]{revtex4}

\usepackage{graphicx}
\usepackage{dcolumn}
\usepackage{bm}
\usepackage{amsmath, amssymb}

\begin{document}

\title{Van der Waals CrSBr alloys with tunable magnetic and optical properties}

\author{Shalini Badola}
\affiliation{LNCMI, UPR 3228, CNRS, EMFL, Universit\'e Grenoble Alpes, 38000 Grenoble, France}
\author{Amit Pawbake}
\affiliation{LNCMI, UPR 3228, CNRS, EMFL, Universit\'e Grenoble Alpes, 38000 Grenoble, France}
\author{Bing Wu}
\affiliation{Department of Inorganic Chemistry, University of Chemistry and Technology Prague, Technicka 5, 166 28 Prague 6, Czech Republic}
\author{Aljoscha S\"{o}ll}
\affiliation{Department of Inorganic Chemistry, University of Chemistry and Technology Prague, Technicka 5, 166 28 Prague 6, Czech Republic}
\author{Zdenek Sofer}
\affiliation{Department of Inorganic Chemistry, University of Chemistry and Technology Prague, Technicka 5, 166 28 Prague 6, Czech Republic}
\author{Rolf Heid}
\affiliation{Institute for Quantum Materials and Technologies, Karlsruhe Institute of Technology, 76131 Karlsruhe, Germany}
\author{Clement Faugeras}
\email{clement.faugeras@lncmi.cnrs.fr} \affiliation{LNCMI, UPR 3228, CNRS, EMFL, Universit\'e Grenoble Alpes, 38000 Grenoble, France}

\date{\today}

\begin{abstract}
Van der Waals magnets are attracting a lot of attention for their potential integration in spintronic and magnonic technologies. CrSBr is an A-type antiferromagnet that shows a coupling between its electronic band structure and magnetic properties. This property is appealing for applications and it also offers the possibility to investigate magnetic ground states and GHz magnons using visible optics techniques. Using Raman scattering and (magneto)-optical experiments, we describe the magnetic and optical properties of alloys of CrSBr$_{(1-x)}$Cl$_x$ with $x<0.46$. Similar to CrSBr, these alloys are direct band gap semiconductors with a coupling of their electronic and magnetic properties. Exciton energies evolve weakly with composition and we describe the large changes in the saturation magnetic fields and their implications on the magnetic properties. We show that both the interlayer magnetic exchange and electronic interactions are modified by the halogen mixing, offering the possibility to tune magnon energies with alloy composition.
\end{abstract}

\maketitle

Van der Waals (vdW) magnets offer a new breath to solid state magnetism by providing an interesting platform to implement spintronic or magnonic schemes with original opportunities in terms of integration into vdW heterostructures~\cite{Sierra2021}. The main reason is the large tunability of the magnetic properties of vdW magnets by external means. These materials present a very rich collection of magnetic ground states and, because of the weak vdW interlayer interaction, they can be thinned down to the two-dimensional limit of a magnetic monolayer~\cite{Lee2016,Huang2017,Liu2018,Long2020,Lu2020,Son2021}. The magnetic properties of such thin layers differ from those of the bulk materials. Lowering their thickness allows inserting these thin layers in field effect structures for instance and their magnetic properties can be modified by an external electric field~\cite{Huang2020i}. Finally, the interlayer distance and all interactions effective across the vdW gap can be efficiently tuned by applying uniaxial strain~\cite{Cenker2020}, hydrostatic pressure~\cite{Pawbake2022,Pawbake2023i}, or by intercalation of atoms or molecules~\cite{Long2023,Feuer2025}.

Another route to tune their properties is to control the material composition and substituting part of the magnetic atoms by others, as was demonstrated for instance in MPX$_3$ compounds~\cite{Rao1992,Goossens2013,Chand1996,Graham2020,Man2000}. It was shown recently that a continuous tuning of the magnon gap amplitude could be achieved in the case of Fe$_x$Ni$_{1-x}$PS$_3$~\cite{LeMardel2024}. Despite the natural alloy disorder, long range magnetic ordering persists in such alloys for any amount of Ni ions inserted in FePS$_3$. Alloying can also concern non magnetic elements such as ligands to modify structural parameters (chemical pressure) or interlayer exchange interactions. Alloys of CrSBr$_{(1-x)}$Cl$_x$ have been investigated from the viewpoint of their magnetization properties~\cite{Telford2023} and the gradual substitution of Br atoms by Cl was shown to affect the magnetic properties. In particular, the magnetic order temperature was shown to decrease significantly when adding Cl and the saturation magnetic fields measured by magnetization were also shown to decrease.

Chromium sulfure bromide (CrSBr) is an air-stable layered direct band gap vdW semiconductor with a band gap close to $1.5$~eV and with a strong optical emission that arises from tightly bound excitons at $1.3$~eV. Below $T_N=132$~K, CrSBr is an A-type antiferromagnet (A-AFM) and is the prototype vdW magnet with a strong coupling between its magnetic properties and its electronic band structure: due to spin dependent interlayer hopping, the exciton energy reflects the magnetic state of the material~\cite{Wilson2021}. In the AFM state, excitons are strongly localized in the individual layers and when applying an external magnetic field, individual spins tend to align along this field and the relative spin arrangement between the layers gradually allows interlayer tunnelling. Excitons become delocalized over the different layers, producing a small decrease of their emission energy that can be measured experimentally. This strong coupling between magnetic and electronic properties, unique to CrSBr, allows for optical excitation and manipulation of magnons~\cite{Bae2022,Diederich2022} and to investigate the magnetic properties of CrSBr by optical means~\cite{Wilson2021,Pawbake2023,Pawbake2023i,komar2024}.

In this Letter, we present an experimental investigation of the optical and magnetic properties of the mixed-halogen compounds CrSBr$_{(1-x)}$Cl$_x$ with $x < 46\%$. As Br or Cl atoms are located in the outer part of the layers, they lie in the van der Waals gaps. Hence making the substitution of Br by Cl atoms is expected to change the interlayer interaction and Cr-Cr super-exchange paths. Using low temperature Raman scattering and optical spectroscopy, we characterize the vibrational properties of the different alloys by the evolution of their phonon spectrum and we determine the exciton energies. We show that these alloys are direct band gap semiconductors with a bright photoluminescence (PL) close to $1.3$~eV, composed of a zero-phonon emission line with phonon replicas at lower energy. Applying an external magnetic field induces changes of the exciton emission energy related to the magnetic state of the Cr ions. We show that the peculiar physics relating the magnetic and electronic properties in CrSBr also exists in these alloys. Both the critical magnetic field values and the single ion anisotropy decrease when incorporating Cl ions. These experiments establish alloys of CrSBr$_{(1-x)}$Cl$_x$ as CrSBr-like systems with similar magneto-optical effects and with the possibility to tune their magnetic properties by adjusting the Cl content.

\begin{figure}[]
\begin{center}
\includegraphics[width=1\linewidth,angle=0,clip]{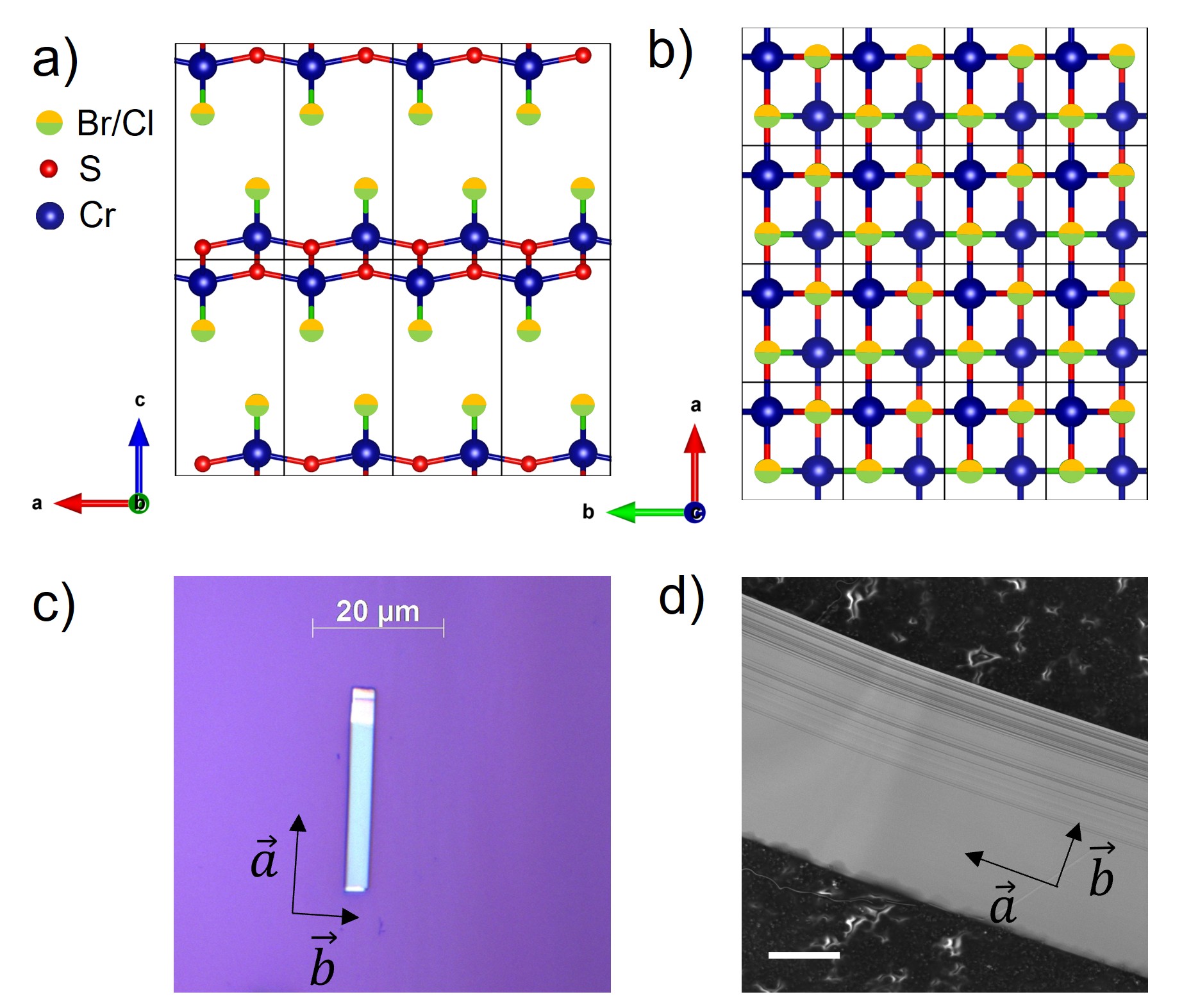}
\end{center}
\caption{a,b) crystal structure of CrSBr$_{(1-x)}$Cl$_x$ alloys viewed along the \textbf{b} and \textbf{c} axis, respectively. Adapted from an original figure prepared using VESTA software~\cite{Momma2011} c) Optical image of an exfoliated flake of CrSBr$_{0.64}$Cl$_{0.36}$ on a SiO$_2$(300 nm)/Si substrate with crystallographic axes. d) Scanning Electron Microscope (SEM) image of CrSBr$_{0.54}$Cl$_{0.46}$ showing the crystallographic axes. The scale bar is $10~\mu m$.}\label{Fig0}
\end{figure}

CrSBr crystallizes in the orthorhombic Pmmn space group, as illustrated in Fig.~\ref{Fig0}a,b. When incorporating Cl atoms, the lattice parameters decrease~\cite{Telford2023} reflecting the different size of Cl and Br atoms, and the crystal structure remains orthorhombic, of FeOCl type. The direct consequence of this crystallographic structure is that, similar to CrSBr, exfoliating CrSBr$_{(1-x)}$Cl$_x$ creates needle-like flakes elongated along the \textbf{a} crystallographic direction which is convenient to identify the crystal axes optically. An optical image of a representative exfoliated flake and a SEM image highlighting the lamellar structure are shown in Fig.~\ref{Fig0}c and \ref{Fig0}d, respectively. All our experiments have been performed on such exfoliated thick (between 70 and 90 nm) flakes. More characterization details of these alloys can be found in the Supplementary Materials.

Figure~\ref{Fig1}a presents characteristic Raman scattering spectra of CrSBr$_{(1-x)}$Cl$_x$ measured at $T = 5$~K. The pure CrSBr compound shows three Ag phonon contributions ($A_g^{1-3}$), typical of orthorhombic systems such as MZX (M a metal ion, Z a chalcogen and X a halogen). Typical phonon line widths in the pure compound are of 0.4~cm$^{-1}$~\cite{Pawbake2023}. These three phonon modes are also found in CrSBr$_{(1-x)}$Cl$_{x}$ alloys with a dispersion of their energies as a function of x, the Cl content, each of the $A_g$ modes showing a different evolution. The energy of $A_g^{1}$ increases at a rate of $0.17~cm^{-1} / \%$ Cl, while for the energy of $A_g^{2}$ decreases at a rate of $-0.23~cm^{-1} / \%$ Cl, and the energy of $A_g^{3}$ increases at a rate of $0.28~cm^{-1} / \%$ Cl. The line width of the three modes increases significantly as a consequence of the alloy disorder that introduces new scattering processes. The phonon line width observed in alloys can be as high as 6-7~cm$^{-1}$. This continuous evolution of the phonon energies when increasing the Cl content suggests that the Cl atoms distribution is homogenous.

\begin{figure}[]
\begin{center}
\includegraphics[width=1\linewidth,angle=0,clip]{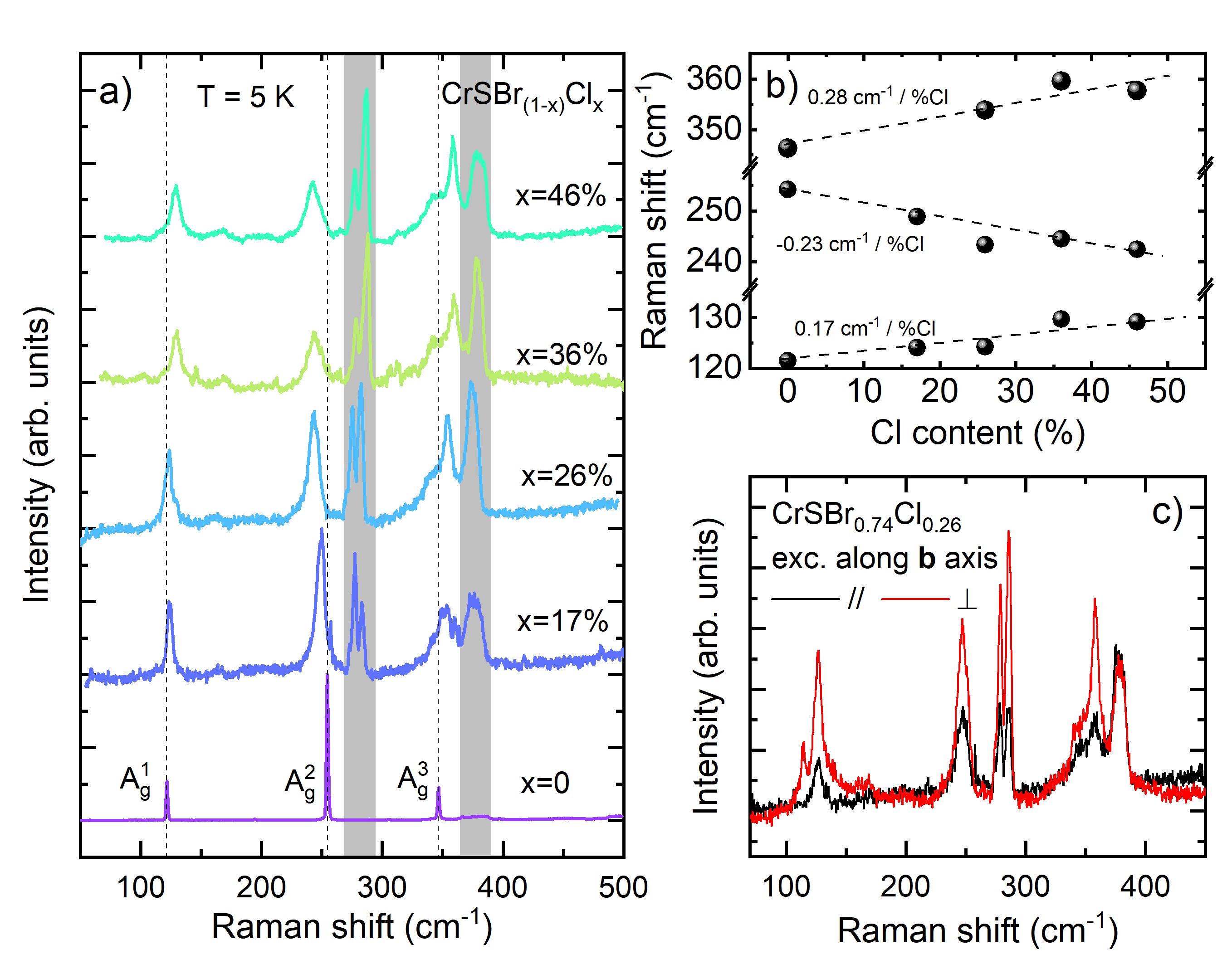}
\end{center}
\caption{a) Low temperature Raman scattering response of CrSBr$_{(1-x)}$Cl$_x$ alloys. b) Energy dispersion of the three $A_g^{1-3}$ modes as a function of the Cl content. c) Polarization resolved Raman scattering spectra of CrSBr$_{0.74}$Cl$_{0.26}$ with the excitation laser polarized along the \textbf{b} axis.}\label{Fig1}
\end{figure}

When incorporating Cl atoms, two new bands appear in the Raman scattering response, within the shaded regions in Fig.~\ref{Fig1}a. They are composed of a doublet of Raman lines at $277~cm^{-1}$ and $283~cm^{-1}$ with an energy increasing by $0.06~cm^{-1} / \%$ Cl and $0.13~cm^{-1} / \%$ Cl, respectively, and of a broad line centered at $375~cm^{-1}$ increasing by $0.11~cm^{-1} / \%$ Cl. The first band shares the same polarization properties as the $A_g$ modes, see Fig.~\ref{Fig1}c. The origin of these two new modes is still uncertain. They could arise from the activation of "silent" or Raman inactive modes due to the disorder introduced by the random Cl atoms but infrared spectroscopy reveals no infrared active mode in this range of energy~\cite{Uykur2025}. From the CrSBr phonon band structure~\cite{Pawbake2023}, we were not able to identify any phonon modes with such energy.

To get insight into the origin of the additional Raman modes observed for Cl substitution of CrSBr, we performed DFT based lattice dynamics calculations. As direct treatment of disorder by ab initio techniques is numerically very demanding, we focused on the special case of 50$\%$ substitution by replacing one Br atom by a Cl atom in the primitive unit cell of the orthorhombic Pmmn structure of CrSBr. This corresponds to a hypothetical compound consisting of ordered Br and Cl layers on each side of the central Cr-S double layer. To take into account the effect of Cl substitution on the lattice constants, we reduced the a, b, and c values of pure CrSBr~\cite{lopez} by 1.6$\%$, 0$\%$, and 3.6$\%$ appropriate for a 50$\%$ substitution~\cite{Telford2023}. Internal atomic positions were then relaxed until the atomic forces were smaller than $2.6\times 10^{-2}$~eV/\AA. The calculational approach closely follows the one adopted in our previous work on bulk CrSBr~\cite{Pawbake2023}, and is briefly outlined in the Supplemental information.

Symmetry classification of the hypothetical structure of CrSBr$_{0.5}$ Cl$_{0.5}$ (space group Pmm2 with point group C$_2v$)) shows that the zone center modes fall into 3 classes: $6 A_1 + 6 B_1 + 6 B_2$. Out of the 6 fully symmetric modes (A$_1$), 1 is an acoustic mode, and 5 are optical modes, all of which involve vibrations of atoms along the c-axis.  Three A$_1$ modes (frequencies 134, 238, and 357 cm$^{-1}$) can be associated to the 3 A$_{1g}$ modes of pure CrSBr. The other two lie at 289 and 368 cm$^{-1}$ and possess larger vibrational contributions from Cl. In particular, the mode at 289 cm$^{-1}$ predominantly involves vibrations of S and Cl only, see Supplemental information. As the space group lacks inversion symmetry, all 18 modes are formally Raman active, but no other mode frequency falls in the range of 250-300 cm$^{-1}$.

For a random Cl substitution, one expects that in certain unit cells both Br atoms are replaced by Cl. To estimate the effect of such a simultaneous replacement on the dynamical properties, we carried out calculations of zone-center phonons for the fully substituted CrSCl compound using the same lattice constants as for the 50$\%$ case. This structure possesses the same space group as CrSBr. Calculated frequencies of the three A$_{1g}$ Raman active modes are 149, 295, and 381 cm$^{-1}$, respectively.  The observed doublet at 277 and 283 cm$^{-1}$ may then be related to the two calculated slightly different mode frequencies of 289 and 295 cm$^{-1}$ when substituting one or both Br atoms in a unit cell by Cl, respectively.

The mode at 375~$cm^{-1}$ does not show any polarization dependence (Fig.~\ref{Fig1}c), whereas such dependence would be expected for a first-order Raman mode. The temperature dependence of the Raman scattering response, see Supplementary materials, also shows that this particular mode is not present at room temperature and only appears below $T\sim100$~K which is also the threshold temperature below which the second order Raman scattering spectrum of pristine CrSBr becomes observable~\cite{Pawbake2023}. Following these arguments, we identify the 375 cm$^{-1}$ Raman scattering feature as part of the second order Raman scattering spectrum.

\begin{figure}[]
\begin{center}
\includegraphics[width=1\linewidth,angle=0,clip]{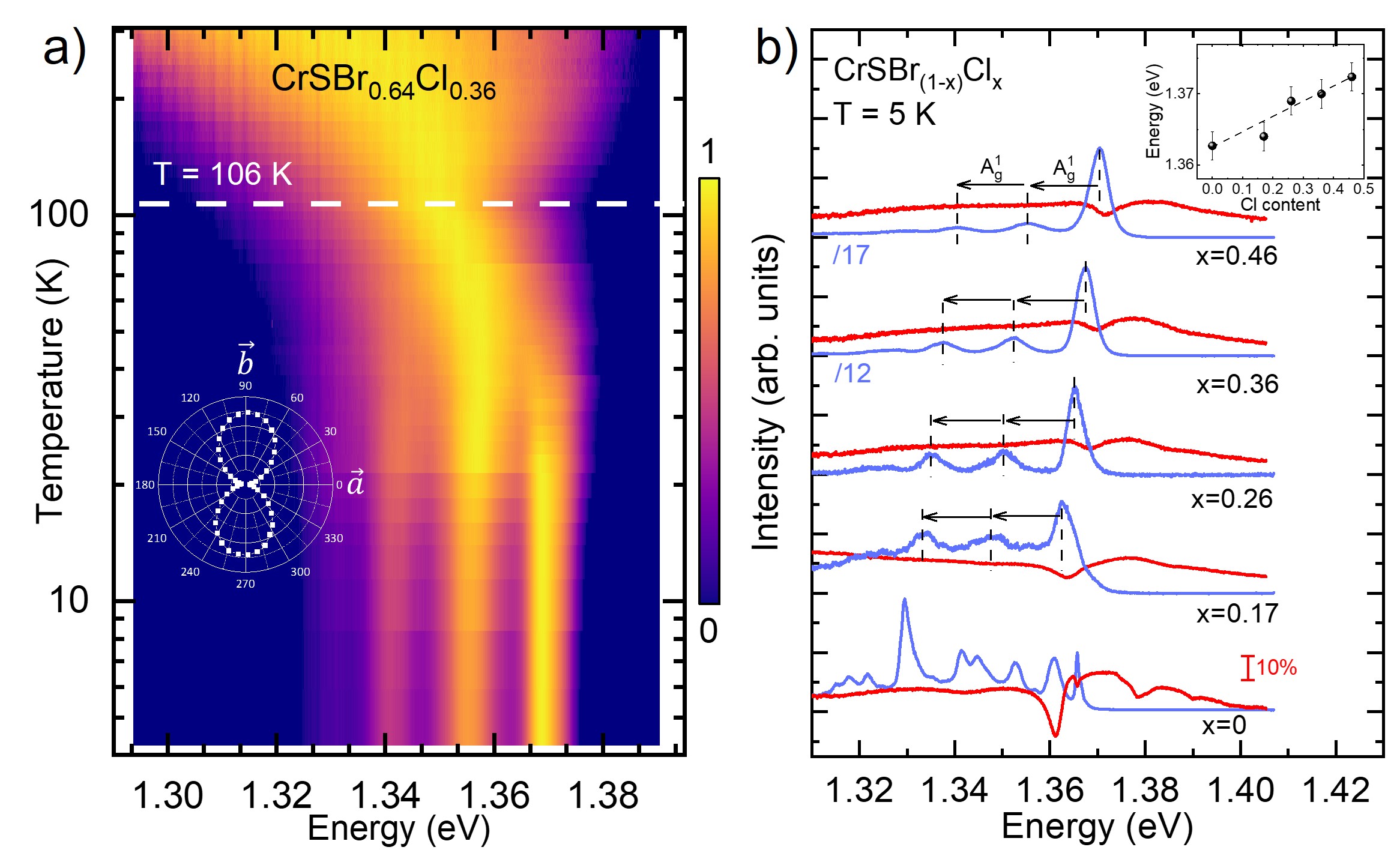}
\end{center}
\caption{a) Temperature evolution of the normalized PL of CrSBr$_{0.64}$Cl$_{0.36}$ and of the reflectance contrast. Inset: PL intensity as a function of polarization angle. b) Low temperature PL spectra of CrSBr$_{(1-x)}$Cl$_x$ alloys (blue solid lines) and reflectivity contrast (red solid lines). Scaling factors used for the normalization of the PL are indicated in blue on the left side of the figure and the horizontal arrows represent a 14 meV energy difference. Inset: Exciton energy as a function of Cl content.}\label{Fig2}
\end{figure}

Similar to pure CrSBr, the mixed halogen alloys show a PL signal at room temperature, see Fig.~\ref{Fig2}a for CrSBr$_{0.64}$Cl$_{0.36}$ and Supplemental information. When decreasing temperature, the broad PL of CrSBr$_{0.64}$Cl$_{0.36}$ becomes narrower down to $T=106$~K. We note that this temperature coincides with the N\'{e}el temperature for this composition identified with magnetization measurements~\cite{Telford2023} and that, similar to the case of the pure compound CrSBr, the critical temperature for the appearance of magnetic order can be determined by the temperature dependence of the PL signal. For lower temperatures, the PL signal transforms for all investigated concentrations into a series of 3 peaks with line widths of $\sim 5$~meV and with a progressively lower intensity at lower energies. Considering layers with thicknesses above $70$~nm (thick layer limit) this line shape does not depend on the thickness. In the inset of Fig.~\ref{Fig2}a, we show that similar to CrSBr, the PL is linearly polarized with a maximum intensity along the \textbf{b} axes reflecting the in-plane anisotropy of optical properties~\cite{Wilson2021}. Figure~\ref{Fig2}b presents the evolution of the low temperature PL and of the reflectance contrast (RC) spectra of CrSBr$_{(1-x)}$Cl$_x$. The reflectance contrast is defined as $RC=(R_s-R_{ref})/(R_s+R_{ref})$ where $R_s$ and $R_{ref}$ are the reflectance of the sample and of the substrate, respectively. First, for all compositions, we observe a bright PL signal suggesting that, similar to pure CrSBr, these alloys (up to x=0.46) are all direct band gap semiconductors with a PL signal originating from excitons. Surprisingly, the integrated intensity of the emission of CrSBr$_{0.54}$Cl$_{0.46}$ is $17$ times higher than the one of pure CrSBr. The exciton can be identified using the RC spectra and coincides with the higher energy emission peak. The exciton energy evolves weakly with the Cl content, linearly increasing from $1.365$~eV for pure CrSBr to $1.372$~eV for $x=0.46$ (slope of $0.23~$meV / $\%$Cl), see inset of Fig.~\ref{Fig2}b. Even though Electron Dispersive X-Ray Spectroscopy (EDS) shows a homogenous distribution of Cl atoms at the $\mu$m-scale, we observe that the exciton emission energy can vary on a flake by $\sim 0.15\%$ from one location to another indicating local strains or nm-scale fluctuations of the Cl concentration, see Supplemental information.

In the case of pure CrSBr, the fine structure aside the exciton emission has been attributed to the self-hybridization of exciton and photon in an optical cavity composed by the thin layer of CrSBr material itself~\cite{Dirnberger2023}. These polariton-like emissions are not evenly spaced in energy, strongly depend on the thickness of the sample defining an optical cavity, and they give rise to resonances in the reflectivity spectrum. This situation appears very different from that observed for CrSBr$_{(1-x)}$Cl$_x$ alloys for which the exciton PL lines are much broader. The more intense line at the highest energy, coincides with the resonance observed in the reflectivity contrast (red curves) and other PL lines, not observed in the RC spectra, are separated by multiples of $14$~meV from the main line. This energy corresponds to the $A_g^1$ phonon as measured in Fig.~\ref{Fig1}a. We hence interpret the emission spectrum of CrSBr$_{(1-x)}$Cl$_x$ alloys as being composed of a high energy line corresponding to the bare exciton, i.e. a zero-phonon line, and of low energy phonon replicas involving the exciton radiative decay with the simultaneous emission of n$A_g^1$ phonons where n is an integer~\cite{Yu2005}. Such phonon replicas have recently been observed in CrSBr flakes with similar thicknesses~\cite{Lin2024,Lin2024a}. In pristine CrSBr, polariton resonances have been observed for samples with thicknesses of few 100s nm, well above thicknesses investigated in this work for which very few resonances are expected~\cite{Dirnberger2023}. Also, the ability to observe polariton related resonances depends on the exciton oscillator strength and on the quality factor of the cavity, both of which could be strongly altered in alloys.

The spin dependent interlayer hopping of charge carriers in bulk CrSBr localizes excitons in individual layers when the magnetic moments in the neighbouring layers are anti-aligned, i.e. in the low temperature A-type antiferromagnetic state of CrSBr~\cite{Wilson2021}. When applying an external field, magnetic moments in the different layers will gradually cant to align with the external field, reducing their relative angle and increasing the interlayer hoping. This leads to the delocalization of excitons over different layers and to a measurable decrease of their emission energy. This phenomenon is today unique to CrSBr and offers the possibility to probe and manipulate magnetic properties by optical means in the near infrared and visible range of energies. In particular, the evolution of the interlayer hopping when applying an external magnetic field, and hence of the relative spin alignment can be traced by investigating the exciton emission energy~\cite{Wilson2021, Pawbake2023i}. This effect has also been used recently to detect magnon excitations by the changes in the dielectric function induced by their propagation~\cite{Bae2022}. Whether this physics also applies to the mixed halogen compounds CrSBr$_{(1-x)}$Cl$_x$ is crucial as this family of materials can provide a new platform of magnetic semiconductors with direct optical band gap and coupled magnetic-electronic properties, with optical and magnetic properties tunable by adjusting the Cl content.

\begin{figure*}[]
\begin{center}
\includegraphics[width=1\linewidth,angle=0,clip]{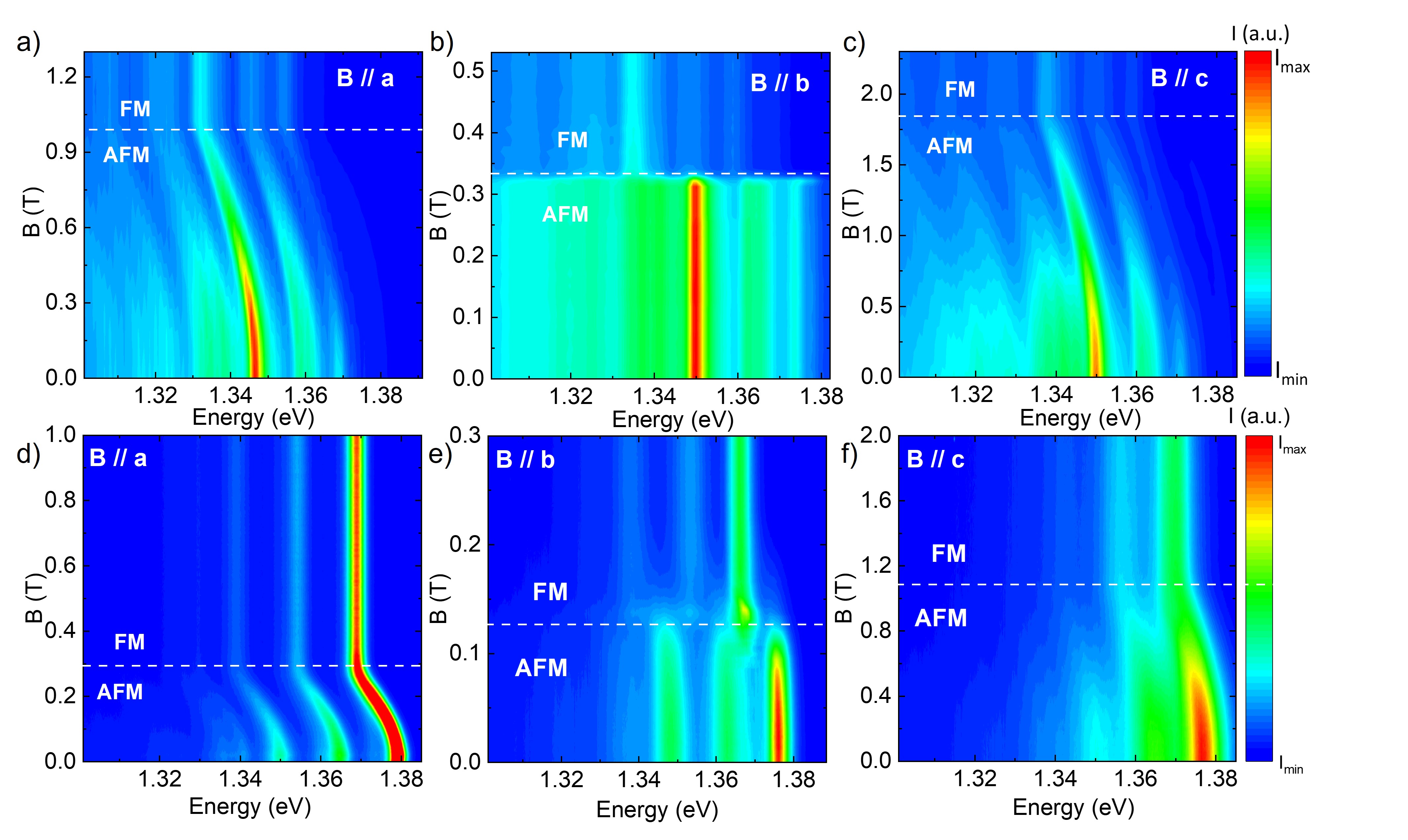}
\end{center}
\caption{Low temperature unpolarized magneto-PL spectra of CrSBr with the magnetic field applied along the \textbf{a}, \textbf{b} anc \textbf{c} axis (a-c) and of CrSBr$_{0.54}$Cl$_{0.46}$ (d-e)).}\label{Fig3}
\end{figure*}

Figure~\ref{Fig3} presents the evolution of the low temperature PL when applying an external magnetic field along the \textbf{a}, \textbf{b}, and \textbf{c} crystallographic axes for the pure CrSBr compound and for CrSBr$_{0.54}$Cl$_{0.46}$. The observed behaviors are identical: along \textbf{a} or \textbf{c} axes, the exciton energy decreases monotonously up to the saturation magnetic field which corresponds to the external field necessary to align all magnetic moments along its direction. Above this saturation magnetic field, all magnetic moments are aligned and excitons are delocalized over many layers. This very specific dependence of the PL energy when changing the external magnetic field shows that similar to bulk CrSBr, CrSBr$_{(1-x)}$Cl$_x$ alloys are direct band gap semiconductors with a magnetic field induced delocalization of excitons due to a spin dependent interlayer hopping of charge carriers. Along the \textbf{a} or \textbf{c} axes, the evolution of the PL energy shows that magnetic moments in the individual layers are canting continuously along the external field direction. Comparing pure CrSBr with the alloy with the CrSBr$_{0.54}$Cl$_{0.46}$ alloy, the values for saturation magnetic fields decrease by 70$\%$ along the \textbf{a} and \textbf{b} axis, and by 40$\%$ along the \textbf{c} axis. Strikingly, when applying the external field along the \textbf{b} axis, both the pure compound and the x=46$\%$ mixed halogen compound show the same abrupt change of the exciton energy, which is typical of an Ising type spin-flip transition~\cite{Cho2023,Pawbake2023i}. This observation sets CrSBr$_{(1-x)}$Cl$_x$ alloys into the category of bi-axial magnetic systems with the easy axis corresponding to the \textbf{b} crystallographic axis. Observing the Ising-type transition also provides information concerning the in-plane magnetic anisotropy as the single-step spin-flip process corresponds to the high anisotropy regime~\cite{Pawbake2023i}. All the investigated alloys showed the Ising-type transition when applying the external magnetic field along the \textbf{b} axis, see Supplemental information.

Figure~\ref{Fig4}a shows the evolution of the saturation magnetic field corresponding to alloy compositions up to 46$\%$ of Cl into CrSBr. These values have been extracted from magneto-PL experiments with the external field applied along the different crystallographic axis and offer an original way, alternative to magnetization measurements~\cite{Telford2023}, to analyze the magnetic properties. The observed trend is a global decrease of the magnetic field necessary to saturate the spin lattice when increasing the content of Cl. This overall decrease is accompanied by a change of the magneto-crystalline anisotropies as values of the saturation field along \textbf{a} axis evolve towards those measured along the \textbf{b} axis, the magnetic easy-axis. This implies that when substituting Br with Cl, the magnetic anisotropies evolve towards those of an easy-plane magnetic system with the easy plane coinciding with the (\textbf{a},\textbf{b}) plane. Despite this similarity in the absolute values of the critical fields, the magneto-optical behavior observed when applying the magnetic field along these two axis remains very distinct, showing a gradual canting of the spins as evidenced by a gradual change of the PL energy for B//\textbf{a} while showing an abrupt Ising-like transition when B//\textbf{b}. The \textbf{c} axis remains the hard axis for the compositions explored in this work.

Considering the model for bi-axial magnetic systems~\cite{Cho2023}, the relations between the critical magnetic fields and the microscopic parameters of the Hamiltonian are :
\begin{eqnarray}
g_a \mu_B B^a_c & = & 2S(4J_\perp + D -E) \,,
\nonumber \\
g_c \mu_B B^c_c & = & 2S(4J_\perp + D +E) \ ,
\nonumber \\
g_b \mu_B B^b_{c} & = & 4J_\perp S \,.
\label{Hb}
\end{eqnarray}
where $g_{\alpha}=2$ are the g-factors when the external magnetic field is applied along the $\mathbf{\alpha}$ axis, ${\bf S}_i$ are $S=3/2$ spins of chromium ions, $J_{\perp}$ is the interlayer exchange coupling constant, $D$ and $E$ are parameters of biaxial single-ion anisotropy with $\alpha$ being either $\mathbf{b}$ (easy axis), $\mathbf{a}$ (intermediate axis ($D>E$)), and $\mathbf{c}$ the hard axis. B$^{\alpha}_c$ are the critical values of magnetic field necessary to align all the magnetic moments along the external field. The intralayer FM exchange interactions are two orders of magnitude larger~\cite{Scheie2022} than the weak interlayer AFM exchange $J_{\perp}$ and forces spins within individual layers to remain parallel to each other.

\begin{figure}[]
\begin{center}
\includegraphics[width=1\linewidth,angle=0,clip]{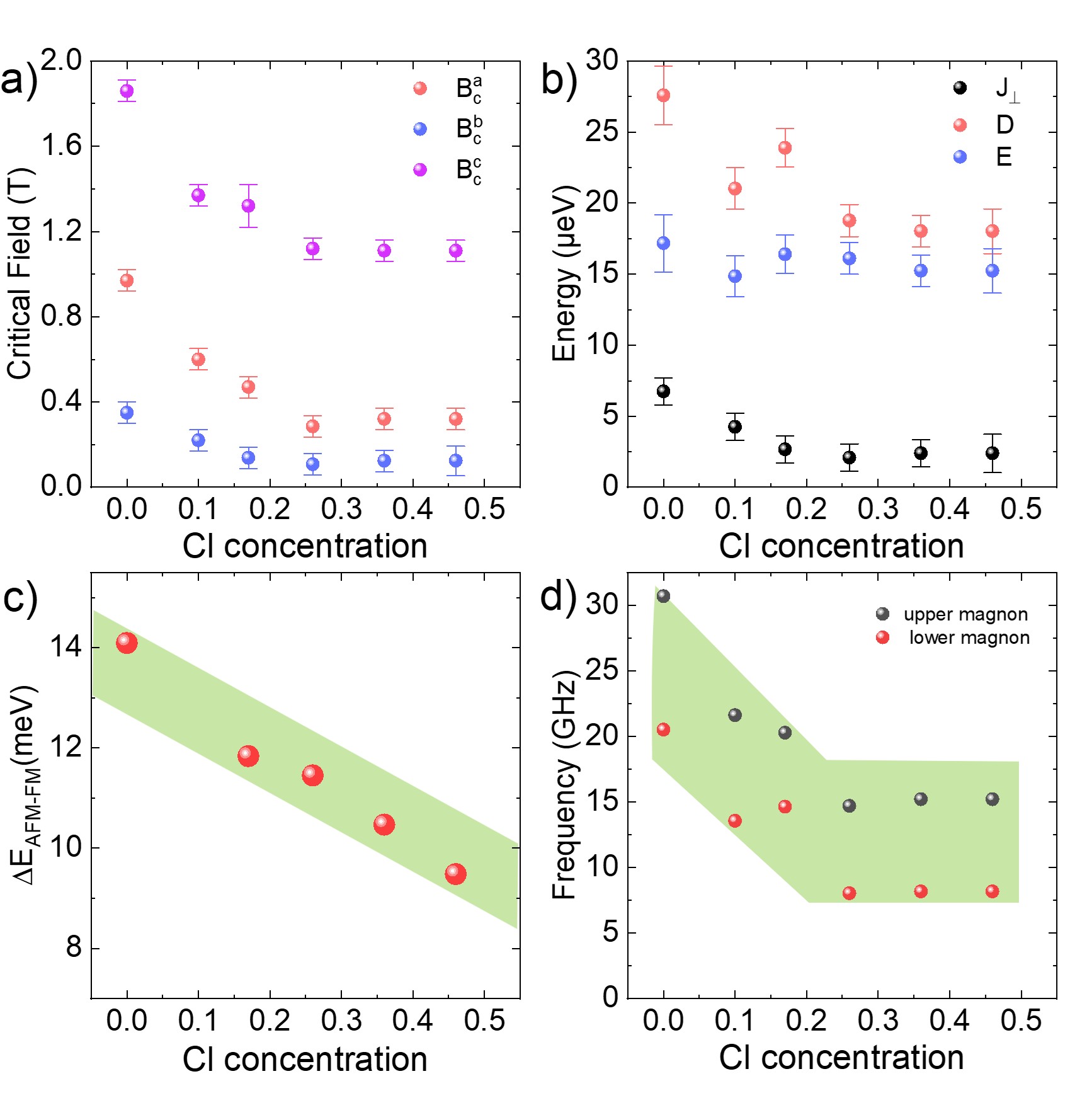}
\end{center}
\caption{a) Evolution of the saturation magnetic fields $B_c^{a,b,c}$ along the three \textbf{a},\textbf{b} and \textbf{c} crystallographic axis. b) microscopic parameters, as a function of the Cl concentration of CrSBr$_{(1-x)}$Cl$_x$ alloys, calculated using the model described in the main text and the values from panel a). c) Average energy difference between exciton in AFM and in FM state as a function of the composition, the green area is a guide for the eyes. d) Calculated magnon gaps using the parameters from panel b), the green area is a guide for the eyes.}\label{Fig4}
\end{figure}

From the knowledge of the saturation magnetic fields and Eq.~\ref{Hb}, we can deduce $J_{\perp}$, D and E for all compositions. Their values are presented in Fig.~\ref{Fig4}b, where their dependence on alloy composition is shown. The interlayer exchange parameter $J_{\perp}$ decreases from 6.7 $\mu eV$ in pure CrSBr down to 2.1 $\mu eV$ for $x>0.3$ and then remains relatively constant up to $x=0.46$. The single ion anisotropy D is also significantly decreasing from 27.5~$\mu eV$ down to 18~$\mu eV$. Strangely, we find that the in-plane anisotropy E is very weakly affected by the Cl substitution, changing from 17~$\mu eV$ to 15~$\mu eV$. From this model, the Cl substitution mainly affects the interlayer exchange $J_{\perp}$ and the single ion anisotropy D.

The coupling between the electronic band structure and the magnetic configuration manifests itself in these experiment as a decrease of the exciton emission energy when the magnetic moments in adjacent layers are aligned. For all the alloys investigated in this work, we observe a decrease of the exciton energy when saturating the magnetic moments along the external field direction, but this energy difference tends to decrease with the Cl content, see Fig.~\ref{Fig4}c. In the picture of excitons localized in individual layers in AFM state and delocalized in the FM state when all spins are aligned, this observation implies a weaker interlayer electrostatic coupling. Substituting Br atoms by Cl atoms changes the interlayer orbitals overlap which translates into two effects: a decrease of the interlayer magnetic exchange interaction as reflected by the pronounced decrease of $J_{\perp}$ shown in Fig.~\ref{Fig4}b, and also by a decreases the electronic interlayer hopping as evidenced by the observed decrease of the exciton energy difference between AFM and FM states.

The evolution of microscopic parameters presented in Fig.~\ref{Fig4}b is of prime importance as these parameters also determine the amplitude of the magnon gaps. Bi-axial CrSBr~\cite{Cho2023} presents two magnon gaps defined by:
\begin{equation}
\Delta_{u,l} = 2S\sqrt{(D\pm E)(4J_\perp + D\mp E)} \,,
\label{equ4}
\end{equation}

As our magneto-optical investigations, together with magnetization measurements on similar alloys~\cite{Telford2023}, indicate that CrSBr$_{(1-x)}$Cl$_x$ alloys are also bi-axial magnetic compounds, we use this equation to evaluate the magnon gaps. We find that these gaps fall in the microwave range of energy. Their evolution with the Cl content, calculated from the microscopic parameters and using Eq.~\ref{equ4}, is presented in Fig.~\ref{Fig4}d. The magnon energy is expected to decrease when increasing the Cl content, and the energy separation between the two modes remains constant, reflecting the fact that the in-plane anisotropy E varies weakly with the Cl content. The magnon gaps can therefore be tuned by varying the Cl concentration, which directly modifies the interlayer magnetic exchange interaction.

In conclusion, our comprehensive optical and magneto-optical study demonstrates that mixed halogen alloys CrSBr$_{1-x}$Cl$_x$ retain the essential features of pristine CrSBr while providing an effective route to tune their properties. Raman scattering, photoluminescence, and magneto-photoluminescence reveal that the direct semiconducting gap and the strong exciton–spin coupling characteristic of CrSBr are preserved across the alloy series. At the same time, the substitution of Br by Cl reduces the interlayer exchange interaction and the single-ion anisotropy, thereby offering a direct means to control magnon energies and magnetic anisotropies.
These findings establish CrSBr$_{1-x}$Cl$_x$ as a versatile platform to investigate collective excitations in layered magnetic semiconductors. The ability to modulate both excitonic and magnetic properties by alloying introduces an additional degree of freedom that can be exploited to probe the interplay between magnons, excitons, and phonons in a highly controllable fashion. Beyond the fundamental significance, such tunability opens opportunities for engineering exciton–magnon interactions for opto-spintronics and magnonics.

\begin{acknowledgements}

We acknowledge discussions with J. Coraux, P. Kossacki, M.E. Zhitomirsky and M. Potemski. This work was supported by ANR-23-QUAC-0004, by CEFIPRA project 7104-2 and by LNCMI, member of the European Magnetic Field Laboratory (EMFL). Z.S. was supported by project LUAUS25268 from Ministry of Education Youth and Sports (MEYS) and by the project Advanced Functional Nanorobots (reg. No. CZ.$02.1.01/0.0/0.0/15_003/0000444$ financed by the EFRR). R.H. acknowledges support by the state of Baden-W\"{u}rttemberg through bwHPC.
\end{acknowledgements}


\end{document}